\documentclass[11pt, a4paper]{article}
\usepackage{latexsym}
\usepackage{graphicx}
\usepackage{cite}
\usepackage{amsmath}
\usepackage{amssymb}

\input{tcilatex}

\begin{document}

\title{ Investigating Auto-correlation of Scattered Light of Mixed Particles}
\author{Yong Sun{$\thanks{%
Email: ysun2002h@yahoo.com.cn}$}}
\maketitle

\begin{abstract}
In this work, the normalized time auto-correlation function of the electric
field of the light $g^{\left( 1\right) }\left( \tau \right) $ that is
scattered by the two kinds of particles in dispersion is investigated. The
results show that the logarithm of $g^{\left( 1\right) }\left( \tau \right) $
can be consistent with a line and many reasons can cause the deviations
between an exponentiality and plots of $g^{\left( 1\right) }\left( \tau
\right) $ as a function of delay time $\tau $. The nonexponentiality of $%
g^{\left( 1\right) }\left( \tau \right) $ is not only determined by the
particle size distribution and scattering angle but also greatly influenced
by the relationship between the concentrations, mass densities and the
values that the refractive index of the material expands as a function of
the concentration of the two kinds of particles.
\end{abstract}

\section{INTRODUCTION}

For colloidal dispersion systems, light scattering is a widely used
technique to measure the sizes of particles. The dynamic light scattering
(DLS) technique is to measure the particle sizes from the normalized time
auto-correlation function of the scattered light $g^{\left( 2\right) }\left(
\tau \right)$. The cumulants\cite{re1,re2,re3,re4} has been used as a
standard method to obtain the ploy-dispersity of particles from the DLS
data. In general, when the DLS data are analyzed, the size distribution of
particles is considered as a mono-disperse distribution if the logarithm of
the normalized time auto-correlation function of the electric field of the
light $g^{\left( 1\right) }\left( \tau \right)$ is consistent with a line.
The poly-dispersity of the particles is obtained from the deviation between
an exponentiality and $g^{\left( 1\right) }\left( \tau \right)$.

In the previous work\cite{re5,re6}, for the particles that the number
distribution is Gaussian, it has been discussed that the nonexponentiality
of $g^{\left( 2\right) }\left( \tau \right)$ is determined by the particle
size distribution and scattering angle. In general, the effects of the
particle size distribution are small on the deviation between an
exponentiality and $g^{\left( 2\right) }\left( \tau \right)$ and very large
on the initial slope of the logarithm of $g^{\left( 2\right) }\left( \tau
\right)$ and the effects of the scattering angle are determined by the
particle size distribution and mean particle size. Under some conditions,
the nonexponentiality of $g^{\left( 2\right) }\left( \tau \right)$ is
greatly influenced by the scattering angle.

In this work, the normalized time auto-correlation function of the electric
field of the light $g^{\left( 1\right) }\left( \tau \right) $ that is
scattered by two kinds of particles in dispersion is investigated. The
results show that the logarithm of $g^{\left( 1\right) }\left( \tau \right) $
can be consistent with a line and many reasons can cause the deviations
between an exponentiality and plots of $g^{\left( 1\right) }\left( \tau
\right) $ as a function of delay time $\tau $. The nonexponentiality of $%
g^{\left( 1\right) }\left( \tau \right) $ is not only determined by the
particle size distribution and scattering angle but also greatly influenced
by the relationship between the concentrations, mass densities and the
values that the refractive index of the material expands as a function of
the concentration of the two kinds of particles in dispersion.

\section{THEORY}

For the two kinds of dilute poly-disperse homogeneous spherical particles in
dispersion where the Rayleigh-Gans-Debye approximation is valid, the total
normalized time auto-correlation function of the electric field of the
scattered light $g^{\left( 1\right) }\left( \tau \right) $ can be written as

\begin{equation}
g_{total}^{\left( 1\right) }\left( \tau \right) =\frac{g_{1}^{\left(
1\right) }\left( \tau \right) +\frac{I_{s2}}{I_{s1}}g_{2}^{\left( 1\right)
}\left( \tau \right) }{1+\frac{I_{s2}}{I_{s1}}} ,  \label{Gtotal}
\end{equation}
where $I_{s1}$, $I_{s2}$, $g^{\left( 1\right) }_1\left( \tau \right)$ and $%
g^{\left( 1\right) }_2\left( \tau \right)$ can be obtained as\cite{re7}

\begin{equation}
\frac{I_{s1}}{I_{inc}}=\frac{4\pi ^{2}\sin ^{2}\vartheta n_{s}^{2}\left( 
\frac{dn_1}{dc_1}\right) _{c_1=0}^{2}c_1}{\lambda ^{4}r^{2}}\frac{4\pi
\rho_1 }{3} \frac{\int_{0}^{\infty }R_{s1}^{6}P\left( q,R_{s1}\right)
G\left( R_{s1}\right) dR_{s1}}{\int_{0}^{\infty }R_{s1}^{3}G\left(
R_{s1}\right) dR_{s1}},  \label{Ione}
\end{equation}

\begin{equation}
\frac{I_{s2}}{I_{inc}}=\frac{4\pi ^{2}\sin ^{2}\vartheta n_{s}^{2}\left( 
\frac{dn_2}{dc_2}\right) _{c_2=0}^{2}c_2}{\lambda ^{4}r^{2}}\frac{4\pi
\rho_2 }{3} \frac{\int_{0}^{\infty }R_{s2}^{6}P\left( q,R_{s2}\right)
G\left( R_{s2}\right) dR_{s2}}{\int_{0}^{\infty }R_{s2}^{3}G\left(
R_{s2}\right) dR_{s2}},  \label{Itwo}
\end{equation}

\begin{equation}
g^{\left( 1\right) }_1\left( \tau \right) =\frac{\int R_{s1}^{6} P\left(
q,R_{s1}\right)G\left( R_{s1}\right) \exp \left( -q^{2}D_1\tau \right)
dR_{s1}}{\int R_{s1}^{6}P\left( q,R_{s1}\right) G\left( R_{s1}\right) dR_{s1}%
},  \label{Gone}
\end{equation}

\begin{equation}
g^{\left( 1\right) }_2\left( \tau \right) =\frac{\int R_{s2}^{6} P\left(
q,R_{s2}\right)G\left( R_{s2}\right) \exp \left( -q^{2}D_2\tau \right)
dR_{s2}}{\int R_{s2}^{6}P\left( q,R_{s2}\right) G\left( R_{s2}\right) dR_{s2}%
},  \label{Gtwo}
\end{equation}
here $\vartheta$ is the angle between the polarization of the incident
electric field and the propagation direction of the scattered field, $c$ is
the mass concentration of particles, $r$ is the distance between the
scattering particle and the point of the intensity measurement, $\rho $ is
the density of the particles, $I_{inc}$ is the incident light intensity, $%
I_{s}$ is the intensity of the scattered light that reaches the detector, $%
R_{s}$ is the static radius of a particle, $\ q=\frac{4\pi }{\lambda }%
n_{s}\sin \frac{\theta }{2}$ is the scattering vector, $\lambda $ is the
wavelength of the incident light in vacuo, $n_{s}$\ is the solvent
refractive index, $\theta $ is the scattering angle, $P\left( q,R_{s}\right) 
$ is the form factor of homogeneous spherical particles

\begin{equation}
P\left( q,R_{s}\right) =\frac{9}{q^{6}R_{s}^{6}}\left( \sin \left(
qR_{s}\right) -qR_{s}\cos \left( qR_{s}\right) \right) ^{2}  \label{P(qr)}
\end{equation}
and $G\left( R_{s}\right) $ is the number distribution of particle sizes. In
this work, the number distribution is chosen as a Gaussian distribution

\begin{equation}
G\left( R_{s};\left\langle R_{s}\right\rangle ,\sigma \right) =\frac{1}{
\sigma \sqrt{2\pi }}\exp \left( -\frac{1}{2}\left( \frac{R_{s}-\left\langle
R_{s}\right\rangle }{\sigma }\right) ^{2}\right) ,
\end{equation}
where $\left\langle R_{s}\right\rangle $ is the mean static radius and $%
\sigma $ is the standard deviation,

\noindent From the Einstein-Stokes relation, the diffusion $D$ can be
written as

\begin{equation}
D=\frac{k_{B}T}{6\pi \eta _{0}R_{h}},
\end{equation}
where $\eta _{0}$, $k_{B}$, $T$ and $R_{h}$ are the viscosity of the
solvent, Boltzmann's constant, absolute temperature and hydrodynamic radius,
respectively. The subscripts 1 and 2 show the kinds of the particles in
dispersion.

When $q\rightarrow0$, the Z-average diffusion coefficient $\left\langle
D\right\rangle_{z}$ can be written as

\begin{equation}
\left\langle D\right\rangle_{z} =\frac{\frac{\int_{0}^{\infty }R_{s1}^{6} G
\left( R_{s1}\right) D_1 dR_{s1}}{\int_{0}^{\infty }R_{s1}^{6} G \left(
R_{s1}\right) dR_{s1}}+a\frac{\frac{\int_{0}^{\infty }R_{s2}^{6}G\left(
R_{s2}\right) dR_{s2}}{\int_{0}^{\infty }R_{s2}^{3}G\left( R_{s2}\right)
dR_{s2}}}{\frac{\int_{0}^{\infty }R_{s1}^{6}G\left( R_{s1}\right) dR_{s1}}{%
\int_{0}^{\infty }R_{s1}^{3}G\left( R_{s1}\right) dR_{s1}}}\frac{%
\int_{0}^{\infty }R_{s2}^{6}G\left( R_{s2}\right) D_{2}dR_{s2}}{%
\int_{0}^{\infty }R_{s2}^{6}G\left( R_{s2}\right) dR_{s2}}}{1+a\frac{\frac{%
\int_{0}^{\infty }R_{s2}^{6}G\left( R_{s2}\right) dR_{s2}}{\int_{0}^{\infty
}R_{s2}^{3}G\left( R_{s2}\right) dR_{s2}}}{\frac{\int_{0}^{\infty
}R_{s1}^{6}G\left( R_{s1}\right) dR_{s1}}{\int_{0}^{\infty
}R_{s1}^{3}G\left( R_{s1}\right) dR_{s1}}}}
\end{equation}
where $a=\frac{\left( \frac{dn_{2}}{dc_{2}}\right) _{c_{2}=0}^{2}c_{2}\rho
_{2}}{ \left( \frac{dn_{1}}{dc_{1}}\right) _{c_{1}=0}^{2}c_{1}\rho _{1}}$

\section{RESULTS AND DISCUSSION}

In the previous work\cite{re7,re8}, it was shown that the expected values of
the DLS data calculated based on the commercial and static particle size
information are consistent with the experimental data. In order to
investigate the effects of the mixture of two kinds of particles on the
deviation between an exponentiality and $g^{\left( 1\right) }\left( \tau
\right)$ accurately, the values of $g^{\left( 1\right) }\left( \tau \right)$
were produced directly using Eqs. \ref{Gtotal}, \ref{Gone} and \ref{Gtwo},
respectively.

The values of $g_{1}^{\left( 1\right) }\left( \tau \right) $, $g_{2}^{\left(
1\right) }\left( \tau \right) $ and $g_{total}^{\left( 1\right) }\left( \tau
\right) $ were produced using the information: the temperature $T$,
viscosity of the solvent $\eta _{0}$, wavelength of laser light $\lambda $,
refractive index of the water $n_{s}$ and constant $R_{h}/R_{s}$ were set to
300.49K, 0.8479 mPa$\cdot $S, 632.8 nm, 1.332 and 1.1, scattering angle $%
\theta $ was chosen as 30$^{\text{o}}$ and 90$^{\text{o}}$, mean static
radius $\left\langle R_{s}\right\rangle $ was set to 20 nm, 40 nm, 120 nm
and 200 nm and standard deviation $\sigma $ was 5\% or 20\% of the mean
static radius, respectively.

First investigating the simple situation that the two kinds of particles
have same $c$, $\rho $ and $\left( \frac{dn}{dc}\right) _{c=0}$ or $a=1$.
The values of $\ln {\left( g_{1}^{\left( 1\right) }\left( \tau \right)
\right) }$, $\ln {\left( g_{2}^{\left( 1\right) }\left( \tau \right) \right) 
}$ and $\ln {\left( g_{total}^{\left( 1\right) }\left( \tau \right) \right) }
$ produced using $\left\langle R_{s}\right\rangle $ and $\sigma $ 20 nm, 1
nm and 40 nm, 2 nm are shown in Figs. 1a and 1b for scattering angles 30$^{%
\text{o}}$ and 90$^{\text{o}}$, respectively. Both the results show that the
plots of $\ln {\left( g_{total}^{\left( 1\right) }\left( \tau \right)
\right) }$ as a function of delay time $\tau $ are consistent with a line
respectively and the mixture of the two kinds of particles with the narrow
size distributions investigated do not causes the nonexponentiality of $%
g_{total}^{\left( 1\right) }\left( \tau \right) $ at the scattering angles
investigated. Figure 1 also reveals that the values of $g_{total}^{\left(
1\right) }\left( \tau \right) $ almost are determined by the kind of
particles with the larger mean static radius. It agrees with the fact that
most of scattered light comes from the kind of particles with the larger
mean static radius under the condition $a=1$.

In general, the values of concentration $c$, mass density $\rho $ and the
refractive index of the material expands as a function of the concentration $%
\left( \frac{dn}{dc}\right) _{c=0}$ are different when the two kinds of
particles are mixed. The values of $\ln {\left( g_{total}^{\left( 1\right)
}\left( \tau \right) \right) }$ thus were calculated under the conditions $%
a=5$, 10, respectively. In order to investigate the effects of $a$ on $\ln {%
\left( g_{total}^{\left( 1\right) }\left( \tau \right) \right) }$, the
results for $a=1$, 5 and 10 are shown in Figs. 2a and 2b for scattering
angles 30$^{\text{o}}$ and 90$^{\text{o}}$, respectively. The results reveal
that the larger the value of $a$, the larger the deviations between a line
and plots of $\ln {\left( g_{total}^{\left( 1\right) }\left( \tau \right)
\right) }$ as a function of delay time $\tau $. Comparing to Fig. 1, the
values of $\ln {\left( g_{total}^{\left( 1\right) }\left( \tau \right)
\right) }$ obviously affected by the relative scattered light intensity or $%
c $, $\rho $ and $\left( \frac{dn}{dc}\right) _{c=0}$ of the two kinds of
particles in dispersion.

For wide particle size distributions, the effects of the mixture of two
kinds of particles on the plots of $\ln {\left( g_{total}^{\left( 1\right)
}\left( \tau \right) \right) }$ as a function of delay time $\tau $ are
investigated continually. The values of $\ln {\left( g_{1}^{\left( 1\right)
}\left( \tau \right) \right) }$, $\ln {\left( g_{2}^{\left( 1\right) }\left(
\tau \right) \right) }$ and $\ln {\left( g_{total}^{\left( 1\right) }\left(
\tau \right) \right) }$ were produced using $\left\langle R_{s}\right\rangle 
$ and $\sigma $ 20 nm, 4 nm and 40 nm, 8 nm, respectively. As exploring the
narrow particle size distributions, the simple situation $a=1$ is
investigated first. All results are shown in Figs. 3a and 3b for scattering
angles 30$^{\text{o}}$ and 90$^{\text{o}}$, respectively. Figure 3 shows
that the plots of $\ln {\left( g_{total}^{\left( 1\right) }\left( \tau
\right) \right) }$ as a function of delay time $\tau $ are deviated from a
line and the values almost are determined by the kind of particles with the
larger mean static radius. Comparing to the plots of $\ln {\left(
g_{1}^{\left( 1\right) }\left( \tau \right) \right) }$ and $\ln {\left(
g_{2}^{\left( 1\right) }\left( \tau \right) \right) }$, the
nonexponentiality of $g_{total}^{\left( 1\right) }\left( \tau \right) $ is
large.

Figure 3 shows the same results as Fig. 1 that the values of $%
g_{total}^{\left( 1\right) }\left( \tau \right) $ almost are determined by
the kind of particles with the larger mean static radius. In order to
investigate the general situation that the two kinds of particles are mixed,
the values of $\ln {\left( g_{total}^{\left( 1\right) }\left( \tau \right)
\right) }$ were calculated under the conditions $a=5$, 10, respectively. The
results for $a=1$, 5 and 10 are shown in Figs. 4a and 4b for scattering
angles 30$^{\text{o}}$ and 90$^{\text{o}}$, respectively. The results reveal
the same situation as the particles with narrow particle size distribution
that the larger the value of $a$, the larger the deviations between a line
and plots of $\ln {\left( g_{total}^{\left( 1\right) }\left( \tau \right)
\right) }$ as a function of delay time $\tau $. Comparing to Fig. 3, the
values of $\ln {\left( g_{total}^{\left( 1\right) }\left( \tau \right)
\right) }$ obviously affected by the relative scattered light intensity or $%
c $, $\rho $ and $\left( \frac{dn}{dc}\right) _{c=0}$ of the two kinds of
particles.

When the mean static radius is large enough, it is possible that the
scattered intensity of the kind of particles with the smaller mean static
radius is larger than that of the other at some scattering angles and
smaller than that of the other at other scattering angles. As was discussed
above, the values of $\ln {\left( g_{total}^{\left( 1\right) }\left( \tau
\right) \right) }$ are influenced greatly by the relative scattered light
intensity of the two kinds of particles. It is possible that the
relationships among the plots of $\ln {\left( g_{total}^{\left( 1\right)
}\left( \tau \right) \right) }$, $\ln {\left( g_{1}^{\left( 1\right) }\left(
\tau \right) \right) }$ and $\ln {\left( g_{2}^{\left( 1\right) }\left( \tau
\right) \right) }$ at different scattering angles can show the
characteristic that the relative scattered light intensity of the two kinds
of particles changes as a function of scattering angle. The plots of $\ln {%
\left( g_{1}^{\left( 1\right) }\left( \tau \right) \right) }$, $\ln {\left(
g_{2}^{\left( 1\right) }\left( \tau \right) \right) }$ and $\ln {\left(
g_{total}^{\left( 1\right) }\left( \tau \right) \right) }$ produced using $%
\left\langle R_{s}\right\rangle $ and $\sigma $ 120 nm, 6 nm and 200 nm, 10
nm at scattering angles 30$^{\text{o}}$ and 90$^{\text{o}}$ are used to
explore the effects of the change of the relative scattered light intensity
on $g_{total}^{\left( 1\right) }\left( \tau \right) $. The simple situation $%
a=1$ is investigated first. The results for scattering angles 30$^{\text{o}}$
and 90$^{\text{o}}$ are shown in Figs. 5a and 5b, respectively.

Figure 5 shows that the mixture of the two kinds of particles do not cause a
large deviation between a line and plots of $\ln {\left( g_{total}^{\left(
1\right) }\left( \tau \right) \right) }$. Comparing to Figs. 1b and 3b, the
results at a scattering angle of 90$^{\text{o}}$ show a different feature
that the values of $\ln {\left( g_{total}^{\left( 1\right) }\left( \tau
\right) \right) }$ approximate that of $\ln {\left( g^{\left( 1\right)
}\left( \tau \right) \right) }$ of the kind of particles with the smaller
mean static radius. It agrees with the fact that most of scattered light
comes from the kind of particles with the smaller mean static radius at a
scattering angle of 90$^{\text{o}}$ under the condition $a=1$. Next,
considering the characteristics of Fig. 5, for the mixture of the two kinds
of particles $a$ was chosen as 1, 5 and 10 at a scattering angle of 30$^{%
\text{o}}$ to increase the light intensity scattered by the particles with
the smaller mean static radius and 1, 0.2 and 0.1 at a scattering angle of 90%
$^{\text{o}}$ to decrease the light intensity scattered by the particles
with the smaller mean static radius. The values of $\ln {\left(
g_{total}^{\left( 1\right) }\left( \tau \right) \right) }$ at scattering
angles 30$^{\text{o}}$ and 90$^{\text{o}}$ are shown in Figs. 6a and 6b,
respectively. Comparing to Fig. 5, Fig. 6 reveals that the values of $\ln {%
\left( g_{total}^{\left( 1\right) }\left( \tau \right) \right) }$ are
influenced greatly by the values of the relative scattered intensity of the
two kinds of particles.

For wide particle size distributions, for example $\sigma /\left\langle
R_{s}\right\rangle =20\%$, the plots of $\ln {\left( g_{1}^{\left( 1\right)
}\left( \tau \right) \right) }$, $\ln {\left( g_{2}^{\left( 1\right) }\left(
\tau \right) \right) }$ and $\ln {\left( g_{total}^{\left( 1\right) }\left(
\tau \right) \right) }$ as a function of delay time $\tau $ were explored
further. The results of $\ln {\left( g_{1}^{\left( 1\right) }\left( \tau
\right) \right) }$, $\ln {\left( g_{2}^{\left( 1\right) }\left( \tau \right)
\right) }$ and $\ln {\left( g_{total}^{\left( 1\right) }\left( \tau \right)
\right) }$ produced using $\left\langle R_{s}\right\rangle $ and $\sigma $
120 nm, 24 nm and 200 nm, 40 nm at scattering angles 30$^{\text{o}}$ and 90$%
^{\text{o}}$ are shown in Figs. 7a and 7b, respectively. Figure 7 shows the
same feature as Fig. 5. At a scattering angle of 30$^{\text{o}}$ the values
of $\ln {\left( g_{total}^{\left( 1\right) }\left( \tau \right) \right) }$
almost are determined by the kind of particles with the larger mean static
radius and at a scattering angle of 90$^{\text{o}}$ the values approximate
that of $\ln {\left( g^{\left( 1\right) }\left( \tau \right) \right) }$ of
the kind of particles with the smaller mean static radius. Meanwhile Fig. 7
also reveals that many reasons can cause the deviation between a line and
plots of $\ln {\left( g^{\left( 1\right) }\left( \tau \right) \right) }$.
Due to the deviations that cause by the mixture of the two kinds of
particles or the particle size distribution and scattering angle cannot be
distinguished, it is impossible to infer the size distribution of particles
if other information about the particles in dispersion are unknown at a
single scattering angle.

The effects that the different mixtures of the two kinds of particles in
dispersion for the wide size distribution are investigated continually. As
exploring the narrow particle size distribution, the values of $a$ still
were chosen as 1, 5 and 10 at a scattering angle of 30$^{\text{o}}$ to
increase the light intensity scattered by the particles with the smaller
mean static radius and 1, 0.2 and 0.1 at a scattering angle of 90$^{\text{o}}
$ to decrease the light intensity scattered by the particles with the
smaller mean static radius. The values of $\ln {\left( g_{total}^{\left(
1\right) }\left( \tau \right) \right) }$ at scattering angles 30$^{\text{o}}$
and 90$^{\text{o}}$ are shown in Figs. 8a and 8b, respectively. Comparing to
Fig. 7, Fig. 8 reveals that the values of $\ln {\left( g_{total}^{\left(
1\right) }\left( \tau \right) \right) }$ are greatly influenced by the
different mixtures of the two kinds of particles in dispersion.

\section{CONCLUSION}

The nonexponentiality of $g_{total}^{\left( 1\right) }\left( \tau \right) $
is not only determined by the particle size distribution and scattering
angle but also greatly influenced by the relationship between the
concentrations $c$, mass densities $\rho $ and the values that the
refractive index of the material expands as a function of the concentration $%
\left( \frac{dn}{dc}\right) _{c=0}$ of the two kinds of particles in
dispersion. Under some conditions, the plots of $\ln {\left(
g_{total}^{\left( 1\right) }\left( \tau \right) \right) }$ as a function of
delay time $\tau $ are consistent with a line and the mixture of the two
kinds of particles do not causes the nonexponentiality of $g_{total}^{\left(
1\right) }\left( \tau \right) $. At a single scattering angle, the mixture
of the two kinds of particles or the size distribution of one kind of
particles can makes the deviations between an exponentiality and $g^{\left(
1\right) }\left( \tau \right) $, and the nonexponentiality of $g^{\left(
1\right) }\left( \tau \right) $ are influenced greatly by the different
mixtures of the two kinds of particles. Without other information about the
particles in dispersion, it is impossible to infer the size distribution of
particles accurately only based on $g^{\left( 1\right) }\left( \tau \right) $
at a single scattering angle.

Fig. 1. The differences between the lines and plots of $\ln {\left(
g_{total}^{\left( 1\right) }\left( \tau \right) \right) ,}\ln {\left(
g_{1}^{\left( 1\right) }\left( \tau \right) \right) }$ and $\ln {\left(
g_{2}^{\left( 1\right) }\left( \tau \right) \right) }$ as a function of the
delay time $\tau $. The symbols show the calculated values obtained using
Eqs. \ref{Gtotal}, \ref{Gone} and \ref{Gtwo}, and the lines show the linear
fitting to the calculated data respectively. The results for the calculated
data at scattering angles 30$^{\text{o}}$ and $^{\text{o}}$ are shown in a
and b, respectively.

Fig. 2. The differences between the lines and plots of $\ln {\left(
g_{total}^{\left( 1\right) }\left( \tau \right) \right) }$ obtained under
a=1, 5 and 10 as a function of the delay time $\tau $. The symbols show the
calculated values obtained using Eq. \ref{Gtotal} and the lines show the
linear fitting to the calculated data during a delay time range
respectively. The results for the calculated data at scattering angles 30$^{%
\text{o}}$ and 90$^{\text{o}}$ are shown in a and b, respectively.

Fig. 3. The differences between the lines and plots of $\ln {\left(
g_{total}^{\left( 1\right) }\left( \tau \right) \right) ,}\ln {\left(
g_{1}^{\left( 1\right) }\left( \tau \right) \right) }$ and $\ln {\left(
g_{2}^{\left( 1\right) }\left( \tau \right) \right) }$ as a function of the
delay time $\tau $. The symbols show the calculated values obtained using
Eqs. \ref{Gtotal}, \ref{Gone} and \ref{Gtwo}, and the lines show the linear
fitting to the calculated data during a delay time range, respectively. The
results for the calculated data at scattering angles 30$^{\text{o}}$ and 90$%
^{\text{o}}$ are shown in a and b, respectively

Fig. 4. The plots of $\ln {\left( g_{total}^{\left( 1\right) }\left( \tau
\right) \right) }$ obtained under a=1, 5 and 10 as a function of the delay
time $\tau $. The symbols show the calculated values obtained using Eq. \ref%
{Gtotal}, respectively. The results for the calculated data at scattering
angles 30$^{\text{o}}$ and 90$^{\text{o}}$ are shown in a and b,
respectively.

Fig. 5. The differences between the lines and plots of $\ln {\left(
g_{total}^{\left( 1\right) }\left( \tau \right) \right) ,}\ln {\left(
g_{1}^{\left( 1\right) }\left( \tau \right) \right) }$ and $\ln {\left(
g_{2}^{\left( 1\right) }\left( \tau \right) \right) }$ as a function of the
delay time $\tau $. The symbols show the calculated values obtained using
Eqs. \ref{Gtotal}, \ref{Gone} and \ref{Gtwo}, and the lines show the linear
fitting to the calculated data respectively. The results for the calculated
data at scattering angles 30$^{\text{o}}$ and 90$^{\text{o}}$ are shown in a
and b, respectively.

Fig. 6. The differences between the lines and plots of $\ln {\left(
g_{total}^{\left( 1\right) }\left( \tau \right) \right) }$ obtained under
a=1, 5 and 10 at a scattering angle of 30$^{\text{o}}$ and 1, 0.2 and 0.1 at
a scattering angle of 90$^{\text{o}}$ as a function of the delay time $\tau $%
. The symbols show the calculated values obtained using Eq. \ref{Gtotal} and
the lines show the linear fitting to the calculated data during a delay time
range respectively. The results for the calculated data at scattering angles
30$^{\text{o}}$ and 90$^{\text{o}}$ are shown in a and b, respectively.

Fig. 7. The differences between the lines and plots of $\ln {\left(
g_{total}^{\left( 1\right) }\left( \tau \right) \right) ,}\ln {\left(
g_{1}^{\left( 1\right) }\left( \tau \right) \right) }$ and $\ln {\left(
g_{2}^{\left( 1\right) }\left( \tau \right) \right) }$ as a function of the
delay time $\tau $. The symbols show the calculated values obtained using
Eqs. \ref{Gtotal}, \ref{Gone} and \ref{Gtwo}, and the lines show the linear
fitting to the calculated data during a delay time range respectively. The
results for the calculated data at scattering angles 30$^{\text{o}}$ and 90$%
^{\text{o}}$ are shown in a and b, respectively.

Fig. 8. The plots of $\ln {\left( g_{total}^{\left( 1\right) }\left( \tau
\right) \right) }$ obtained under a=1, 5 and 10 at a scattering angle of 30$%
^{\text{o}}$ and 1, 0.2 and 0.1 at a scattering angle of 90$^{\text{o}}$ as
a function of the delay time $\tau $. The symbols show the calculated values
obtained using Eq. \ref{Gtotal} respectively. The results for the calculated
data at scattering angles 30$^{\text{o}}$ and 90$^{\text{o}}$ are shown in a
and b, respectively.


\begin{thebibliography}{9}
\bibitem{re1} D. E. Koppel, J. Chem. Phys. \textbf{57}, 4814(1972).

\bibitem{re2} C. B. Bargeron, J. Chem. Phys. \textbf{61}, 2134(1974).

\bibitem{re3} J. C. Brown, P. N. Pusey and R. Dietz, J. Chem. Phys. \textbf{%
62}, 1136(1975).

\bibitem{re4} B. J. Berne and R. Pecora, \textit{Dynamic Light Scattering}
(Robert E. Krieger Publishing Company, Malabar, Florida, 1990).

\bibitem{re5} Y. Sun arxiv.org/abs/physics/0511161.

\bibitem{re6} Y. Sun arxiv.org/abs/physics/0512009.

\bibitem{re7} Y. Sun arxiv.org/abs/physics/0511159.

\bibitem{re8} Y. Sun arxiv.org/abs/physics/0511160.
\end{thebibliography}
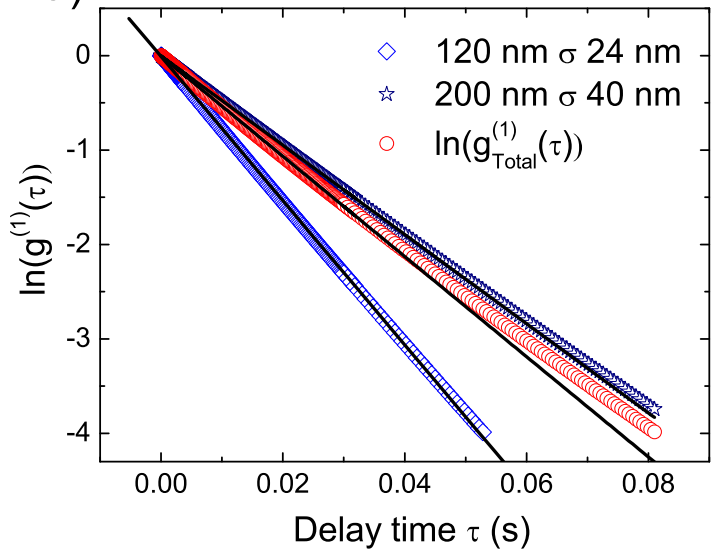
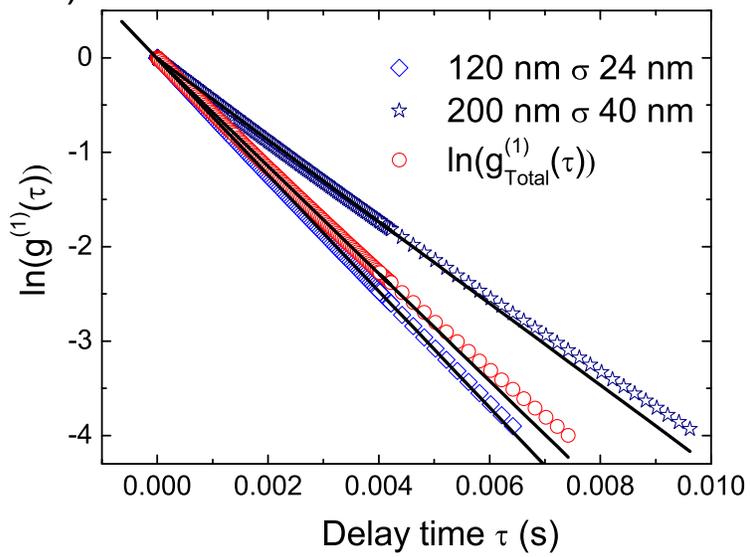
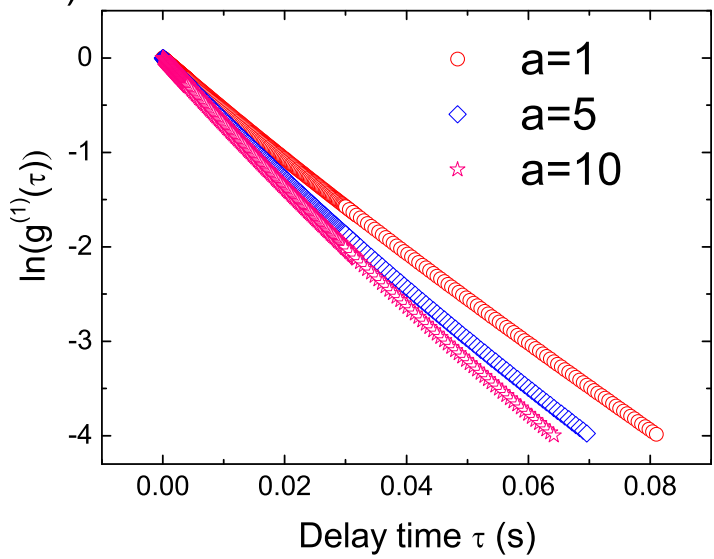
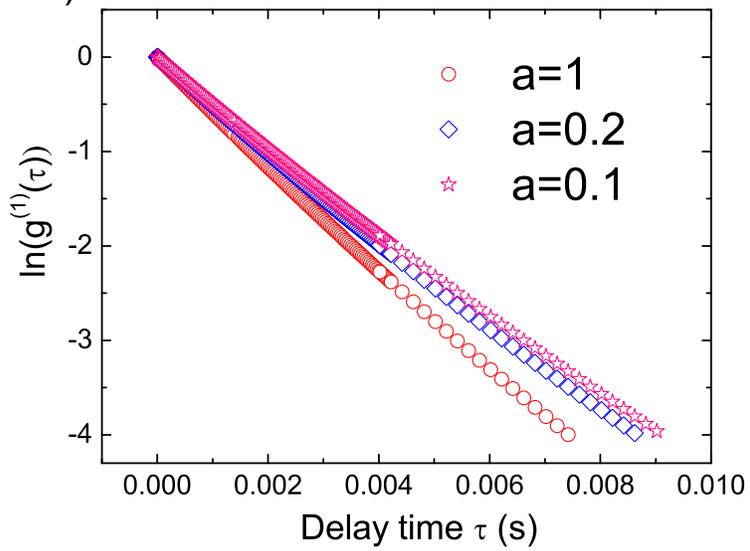
\end{document}